\documentclass[a4paper]{spie}  
\usepackage[]{graphicx}
\usepackage{amssymb,amstext}

\def\farcs{\hbox{$.\!\!^{\prime\prime}$}}

\def\mag{\ifmmode^{\rm m }\else$^{\rm m}$\fi}
\def\fracarcsec{\hbox{$.\!\!^{\rm s}$}}

\def\as{$\,^{\prime\prime}\,$}

\def\hh{\ifmmode^{\rm h}\else$^{\rm h}$\fi}
\def\mm{\ifmmode^{\rm m}\else$^{\rm m}$\fi}
\def\ss{\ifmmode^{\rm s}\else$^{\rm s}$\fi}
\def\deg{\ifmmode^\circ\else$^\circ $\fi}
\def\amin{\ifmmode^\prime\else$^\prime $\fi}
\def\rahm#1#2{\ifmmode{#1}\else{$#1$}\fi\hh\ifmmode{#2}\else{$#2$}\fi\mm}
\def\decdm#1#2{\ifmmode{#1}\else{$#1$}\fi\deg\ifmmode{#2}\else{$#2$}\fi\amin}
\def\ras#1#2{\ifmmode{#1}\else{$#1$}\fi\fracarcsec\ifmmode{#2}\else{$#2$}\fi}
\def\decs#1#2{\ifmmode{#1}\else{$#1$}\fi\farcs\ifmmode{#2}\else{$#2$}\fi}
\def\ra#1#2#3{\ifmmode{#1}\else{$#1$}\fi\hh\ \ifmmode{#2}\else{$#2$}\fi\mm\ \ifmmode{#3}\else{$#3$}\fi\ss}
\def\dec#1#2#3{\ifmmode{#1}\else{$#1$}\fi\deg\ \ifmmode{#2}\else{$#2$}\fi\amin\ \ifmmode{#3}\else{$#3$}\fi\as}

\newcommand{\rab}[4]{\ifmmode{#1}\else{$#1$}\fi\hh\ \ifmmode{#1}\else{$#2$}\fi\mm\ \ifmmode{#1}\else{$#3$}\fi\fracarcsec\ifmmode{#1}\else{$#4$}\fi}
\newcommand{\decb}[4]{\ifmmode{#1}\else{$#1$}\fi\deg\ \ifmmode{#1}\else{$#2$}\fi\amin\ \ifmmode{#1}\else{$#3$}\fi\farcs\ifmmode{#1}\else{$#4$}\fi}

\title{Control interface concepts for CHARA 6-telescope
  fringe tracking with CHAMP+MIRC}

\author{Stefan Kraus\supit{a}, John Monnier\supit{a}, Fabien Baron\supit{a}, Xiao Che\supit{a}, Rafael Millan-Gabet\supit{b}, Nathalie Thureau\supit{c}, Ettore Pedretti\supit{d}
\skiplinehalf
\supit{a}Department of Astronomy, University of Michigan, 500 Church St., Ann Arbor, MI 48109, USA;
\supit{b}California Institute of Technology, NASA Exoplanet Science Institute, Pasadena, CA 91125, USA;
\supit{c}School of Physics and Astronomy, University of St Andrews, North Haugh, St Andrews KY16 9SS, UK;
\supit{d}Scottish Association for Marine Science, Oban, Scotland, UK
}

\authorinfo{Further author information: (Send correspondence to S.K.)\\
E-mail: stefankr@umich.edu, Telephone: +1 734 615 7374}

\begin{document} 
  \maketitle 

\begin{abstract}
Cophasing six telescopes from the CHARA array, the CHARA-Michigan
Phasetracker (CHAMP) and Michigan Infrared Combiner (MIRC) are pushing
the frontiers of infrared long-baseline interferometric imaging in 
key scientific areas such as star- and planet-formation.
Here we review our concepts and recent improvements on the
CHAMP and MIRC control interfaces, which establish 
the communication to the real-time data recording \& fringe tracking code,
provide essential performance diagnostics,
and assist the observer in the alignment and 
flux optimization procedure.
For fringe detection and tracking with MIRC, we have developed 
a novel matrix approach, which provides 
predictions for the fringe positions based on cross-fringe information.
\end{abstract}



\section{INTRODUCTION}
\label{sec:intro}  

The Michigan InfraRed Combiner\cite{mon04,mon10} (MIRC) is a
near-infrared science beam combiner at the
CHARA array located on Mt.\ Wilson in California.  
Using a multiaxial optical layout, MIRC combines the light from all
six beams simultaneously, enabling, for the first time, the
instantaneous measurement of 15 visibility amplitudes and 20 closure
phase triangles in the near-infrared H- and K-band.

In order to enable observations on fainter targets, the CHARA-Michigan
Phasetracker\cite{ber06,ber08,mon10} (CHAMP) uses a co-axial 
pair-wise combination scheme and was optimized for sensitivity by
combining only neighbouring beams.
Using ``phase tracking'' and ``group-delay tracking'' algorithms,
CHAMP can stabilize the atmosphere-induced fringe motion and
provide an enhanced fringe contrast for MIRC and other facility 
instruments.
Together with the uniquely long baselines of the CHARA array (up to
331\,m), MIRC and CHAMP are designed to enable quick-look infrared
long-baseline interferometric imaging of key object classes,
such as the protoplanetary disks around young stars.

In this contribution, we report on our latest developments related to
the user control interface for these instruments. The graphical user
interfaces (GUIs) were implemented in the Python programming language
and evolved from our earlier CHARA observing planning and operational {tools\cite{thu06}}.
They enable an efficient alignment and flux-optimization
(Sect.~\ref{sec:CHAMPalignment}), establish the communication
with the real-time code (Sect.~\ref{sec:CHAMPtracking}),
assist in the fringe search process using cross-fringe information
(Sect.~\ref{sec:MIRCacquisition}),
and allow to initiate the fringe recording (Sect.~\ref{sec:MIRCrecording}).

\subsection{CHAMP alignment}
\label{sec:CHAMPalignment}

\begin{figure}[t]
  \includegraphics[height=17cm,angle=270]{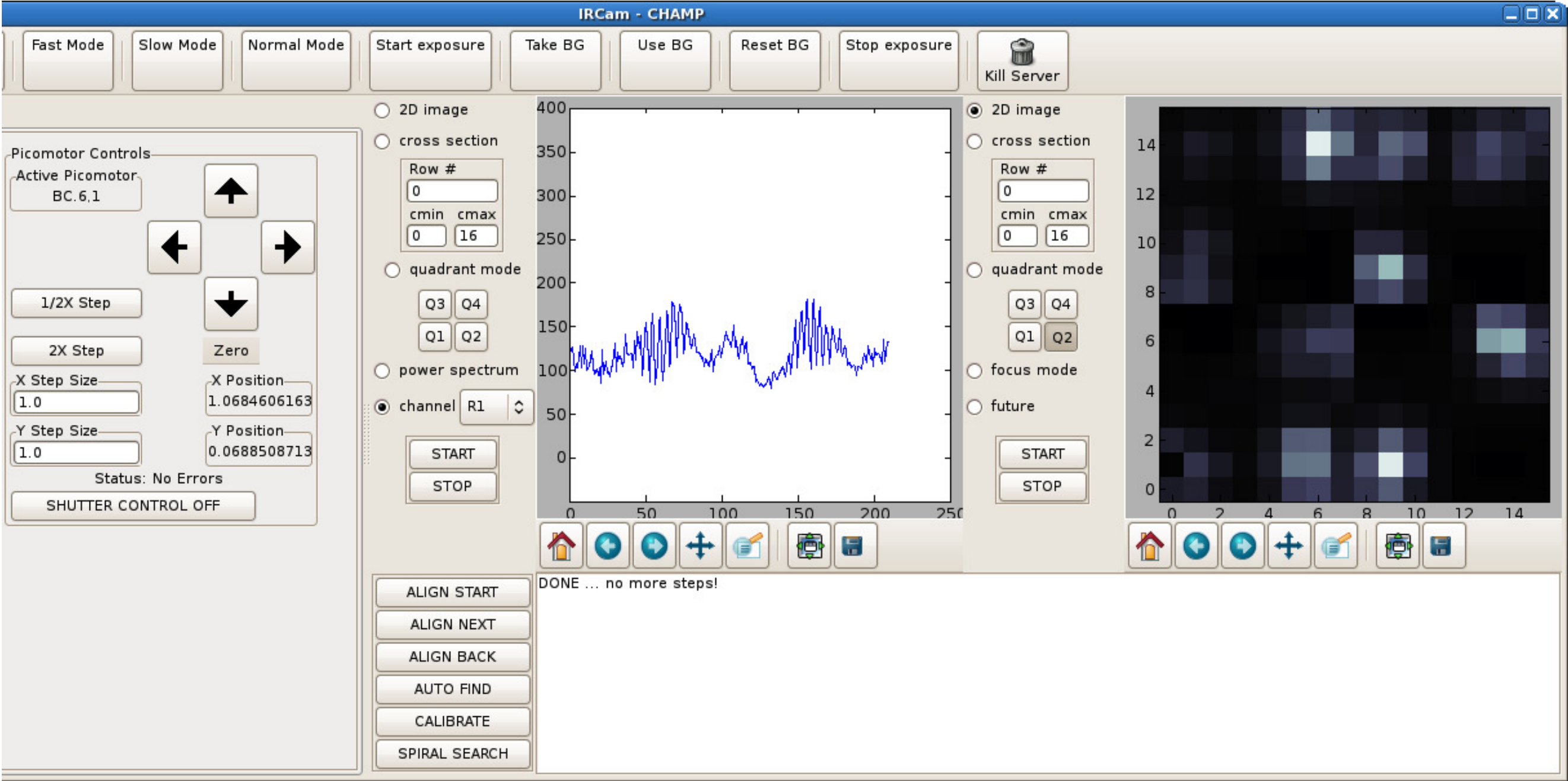}\\
  \caption{The alignment-GUI guides the user through a
    semi-automatical process for flux optimization.}
  \label{fig1}
\end{figure}


Preparing CHAMP for night-time operation requires a careful optimization
of the alignment of all optical components.
For this purpose, we have developed an alignment sequence that consists
of 18 steps, where each step involves the selection of a particular
telescope beam using shutters, followed by the alignment of the
resulting light spot on a reference pixel on the detector.
Besides the large number of alignment steps, the sequence has to be
followed in a precisely defined order. During some steps,
multiple spots are visible at the detector, making it
difficult for the user to identify the correct spot.
As a consequence, the procedure is rather time-consuming and impractical to
perform in a manual fashion, in particular as the alignment procedure
has also to be executed during night-time operation in order to
optimize the flux-injection of faint targets.

Therefore, it was highly desirable to develop a dedicated
alignment-GUI that automates as many steps as possible. 
The interface operates the shutters in order to select a particular
beam, controls the piezo motors to adjust the involved mirrors \& beam
splitters, and displays the real-time detector image.  
Furthermore, the interface outputs written instructions to the user,
and assists him/her in the identification of the spot on the detector,
and shows the appropriate reference pixel onto which the spot has to
be aligned. 

Recently, we have added an auto-alignment feature, which
measures the centroid position of the beam and computes the
piezo motor steps that are required to move the spot onto the reference
pixel. Following a user confirmation, the GUI automatically
operates the actuators in order to perform the alignment.
The procedure is repeated iteratively until convergence is reached.
This automization considerably accelerates the process, yet still gives the
user sufficient control to avoid problems in case a spot has
drifted out of the small field-of-view.

\subsection{CHAMP fringe tracking}
\label{sec:CHAMPtracking}

\begin{figure}[t]
  \includegraphics[height=17cm,angle=270]{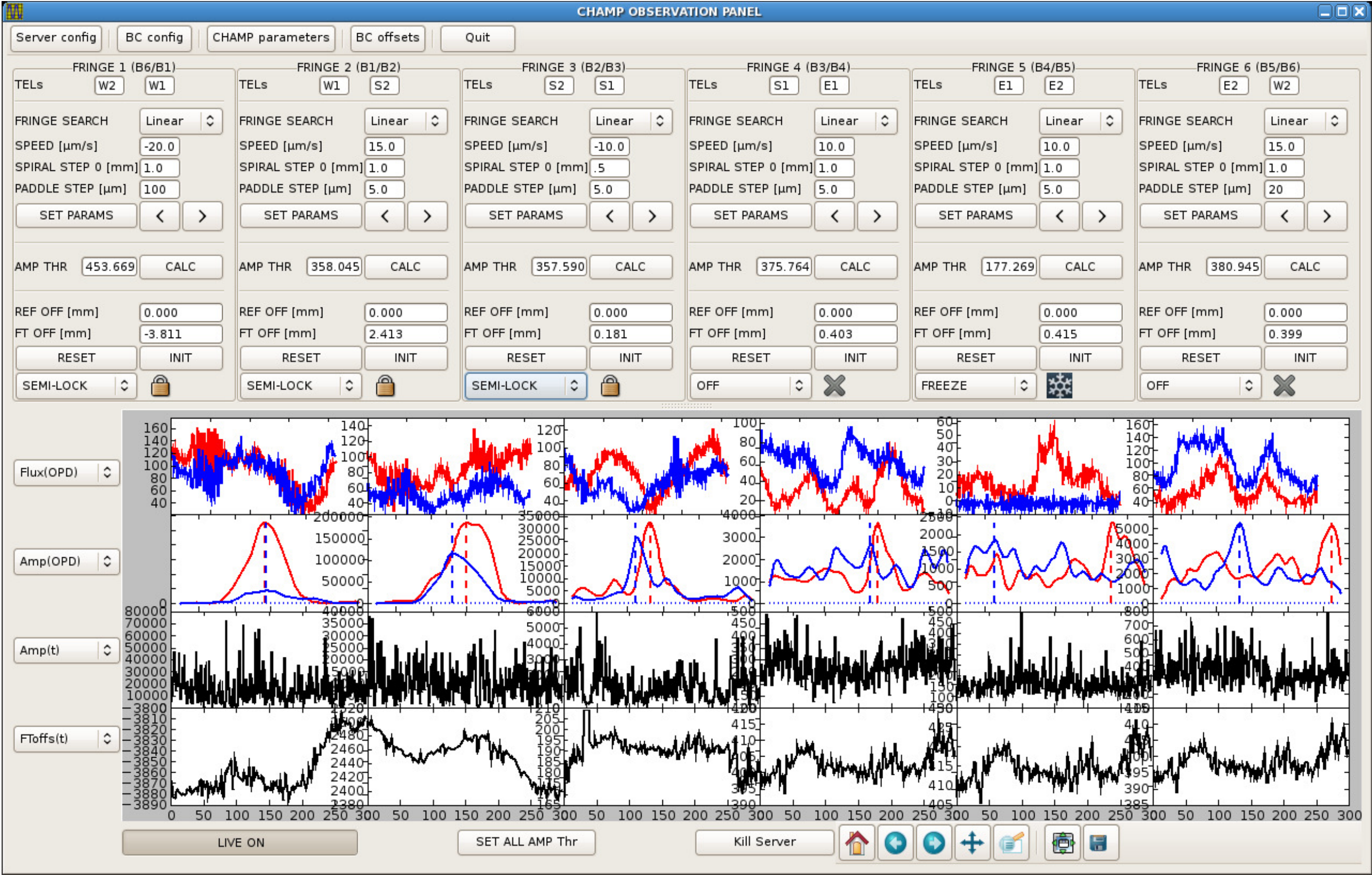}\\
  \caption{Primary GUI for night-time operation of the CHAMP
    6-telescope fringe tracker.  In the shown example, fringes on three of
    the six baselines are locked.}
  \label{fig2}
\end{figure}

In order to enable an efficient operation of CHAMP,
we have developed an interface that allows the user to communicate
with the real-time code and to switch between different fringe search
and fringe tracking states:
\begin{itemize}
\item {\bf SEARCH:} Starts a fringe search, using either a linear or
  spiral search pattern. 
\item {\bf LOCK:} Locks the fringe using a co-phasing algorithm.
\item {\bf SEMI-LOCK:} Locks the fringe using a group-delay tracking
  algorithm.
\item {\bf FREEZE:} Deactivates the fringe tracking, but ensures that
  the delay line position is maintained.
\item {\bf OFF:} Deactivates the fringe tracking, without enforcing
  the current delay position.  Delay lines in this state will be
  used to accumulate delay offsets from other baselines.
\end{itemize}

Besides controlling the tracking state, the GUI provides the user 
with the possibility to move the delay lines manually and to set
the detection threshold for each baseline.
The interface also displays important diagnostic plots, 
such as the measured flux, the power spectrum fringe amplitude, and
the fringe tracker offsets that are sent to the delay lines.
Many parameters can be plotted as function of the optical path delay
(OPD) within an individual scan (e.g.\ ``Flux(OPD)'' in Fig.~\ref{fig2}), or
as a gliding time history plot (e.g.\ ``FToffs(t)'' in Fig.~\ref{fig2}). 
Given the large number of parameters that might be of interest for
diagnostic purposes, we have implemented a dynamical plot allocation,
where the user can select the four parameters that ought to be plotted.
This concept allows us to operate CHAMP using a single graphical
interface.

\subsection{MIRC-6T fringe search \& self-coherencing}
\label{sec:MIRCacquisition}

\begin{figure}[t]
  \includegraphics[height=17cm,angle=270]{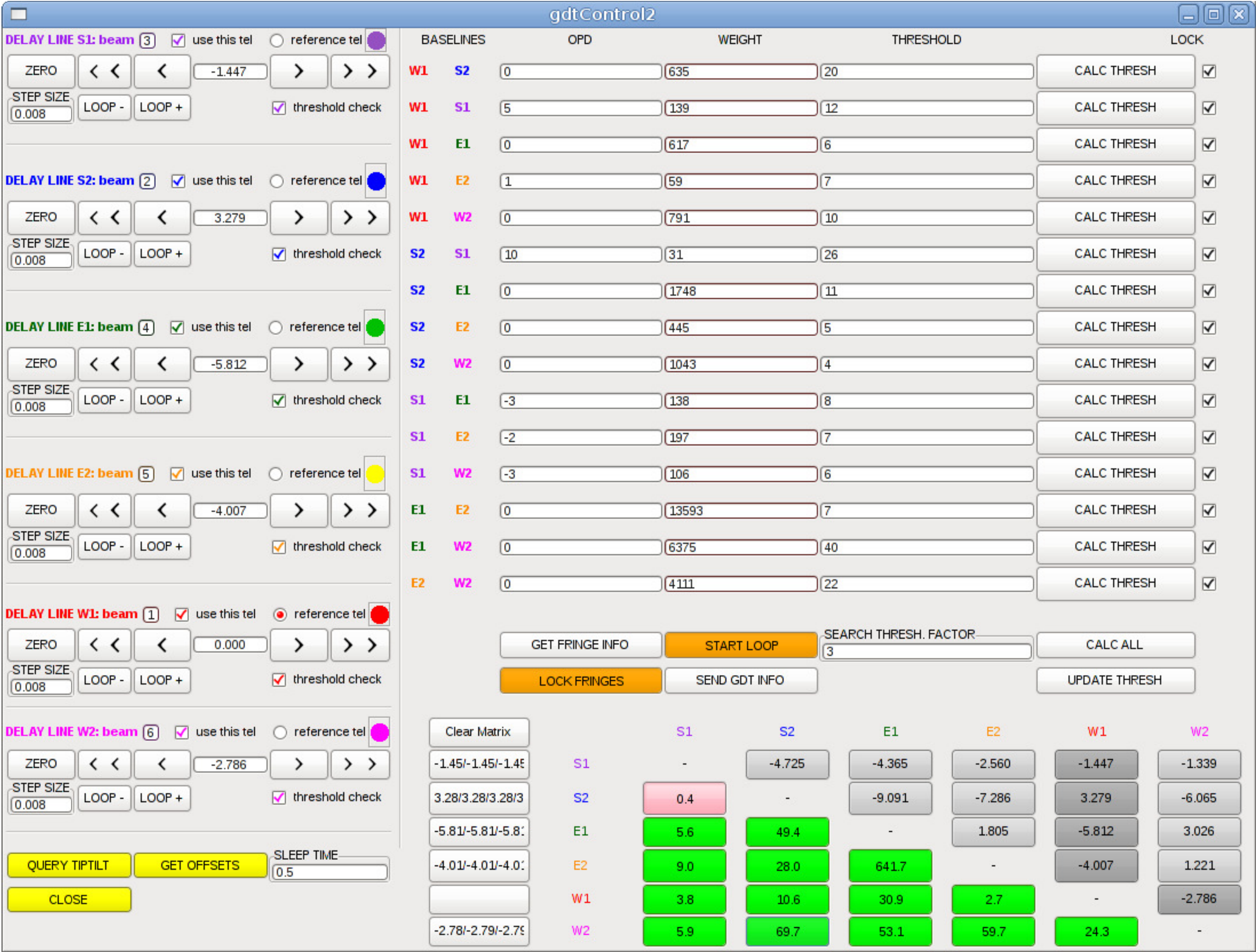}\\
  \caption{Interface enabling fringe search and fringe acquisition
    for the MIRC 6-telescope beam combiner.}
  \label{fig3}
\end{figure}

With its pairwise combination scheme, CHAMP is optimized for
sensitivity, but probes only six of the 15 possible baselines.
This might result in problems for observations on strongly
resolved sources, where baseline bootstrapping becomes essential.
For sufficiently bright objects ($H\gtrsim 6.0$), it is possible in such cases 
to operate MIRC in a self-coherencing mode, where the MIRC
interferograms itself are used for fringe search and group-delay
tracking.

For this purpose, we have developed a dedicated GUI (Fig.~\ref{fig3})
that allows the user to search and track fringes on all 15 baselines.
For each baseline, the user can define a threshold value for fringe 
detection, or ask the software to provide an estimate that
is computed from the background level in the power spectrum 
(buttons ``CALC THRESH'').  
Based on these threshold values, the software computes for each
baseline a signal-to-threshold ratio (STR), which is used both for
fringe search and group-delay tracking.

The user can initiate the fringe search for the individual baselines
using the ``LOOP+'' or ``LOOP-'' buttons.  As soon as a fringe signal
with sufficient STR is detected, the search loop opens and the user
can lock the fringe, either by activating the check marks in the
top-right quadrant, or using the matrix in the bottom-right quadrant
of the control panel (Fig.~\ref{fig3}).
This matrix displays the fringe detection \& tracking information in a
graphical and intuitive way. 
The buttons in the bottom-left quadrant of the matrix output the fringe
STR, where the color of the individual buttons indicates:
\begin{itemize}
\item {\bf grey:} Tracking loop open and no fringe found (STR$<$1).
\item {\bf yellow:} Tracking loop open, but fringe found (STR$\geq$1).
\item {\bf green:} Tracking loop closed and fringe found (STR$\geq$1).
\item {\bf red:} Tracking loop closed, but fringe lost (STR$<$1).
\end{itemize}

In practice, this procedure is used as follows: The user initiates the
fringe search process on five baselines by clicking the
``LOOP+''/``LOOP-'' buttons, while keeping the delay line for the
sixth telescope (``reference telescope'') at a fixed zero position 
(e.g.\ W1 in the example shown in Fig.~\ref{fig3}).
Initially, all buttons in the matrix are colored grey.
As soon as a fringe is detected on one of the 15 baselines, the
corresponding button turns yellow and the search loop for this delay
line opens.  If convinced that the detection is real, the
user clicks the corresponding button in the matrix in
order to close the fringe tracking loop. The algorithm then
tracks the fringe at this baseline, as indicated with
a green color in the matrix.  In case the fringe gets lost, the
corresponding button turns red, informing the user about the
problem.

Besides indicating the fringe STR \& tracking state (in the
bottom-left quadrant of the matrix), the matrix provides information
about the delay line position at which the fringe has been detected. 
This information is output in the top-right quadrant of the matrix
with labels giving the OPD difference between the involved 
delay lines at the time of the fringe detection.
For the five baselines involving the reference telescope (S1-W1,
S2-W1, E1-W1, E2-W1, and W1-W2 in Fig.~\ref{fig3}; colored dark grey),
these OPD differences correspond to the delay line 
positions at which the fringes have been found. 
Detecting these {\it reference-baseline fringes} is particularly
useful, as they allow retrieving the absolute (and not only relative)
delay position of the white-light fringe.

During the fringe search, one also frequently encounters 
{\it cross-fringes} that do not include the reference 
telescope (e.g. S1-S2, S1-E1, E2-W1, etc.).
In order to make use of this relative OPD information, we implemented  
an algorithm that predicts the zero-OPD position for delay lines based
on the measurement of (at least) one reference-baseline fringe, and
the OPD difference of any detected cross-fringes.
For instance, if the fringe on baseline S2-W1 has been detected by
scanning on the delay line associated with telescope S2 (where W1 is
the reference telescope) and the cross-fringe for S2-E1 has been detected, then
it is possible to predict the fringe position for baseline E1-W1, and,
thus, the correct position for the delay line involving telescope E1. 
Oftentimes there exist different cross-fringe combinations that enable
such a prediction, all of which can be utilized.  For instance, in
the given example it is also possible to solve for the delay line
involving telescope E1 using cross-fringe detections on the S2-W1 and
E1-W1 baselines. Solving these equations requires very basic math
(addition of the determined OPD differences), but
is too complicated to be carried out by the observer in real-time.
Therefore, we implemented an algorithm in the MIRC control interface 
that automates this process and solves for all cross-fringe combinations
that lead to predictions of reference-baseline fringe positions. 
This process runs automatically in the background, and the predicted
OPD positions are displayed to the left of the matrix.
By clicking the corresponding button, the user can move the delay line
automatically to the predicted zero-OPD position. 
As input for the computation, the algorithm uses the OPD differences
displayed in the upper-right part of the matrix.  The user can
influence these input parameters, for instance by rejecting false
fringe detections, which can be reset by clicking on the button with
the wrong OPD difference values.

This concept accelerates the fringe search process considerably and
becomes more and more effective, the larger the number of combined
apertures.

\subsection{MIRC-6T data recording}
\label{sec:MIRCrecording}

\begin{figure}[t]
  \includegraphics[height=17cm,angle=270]{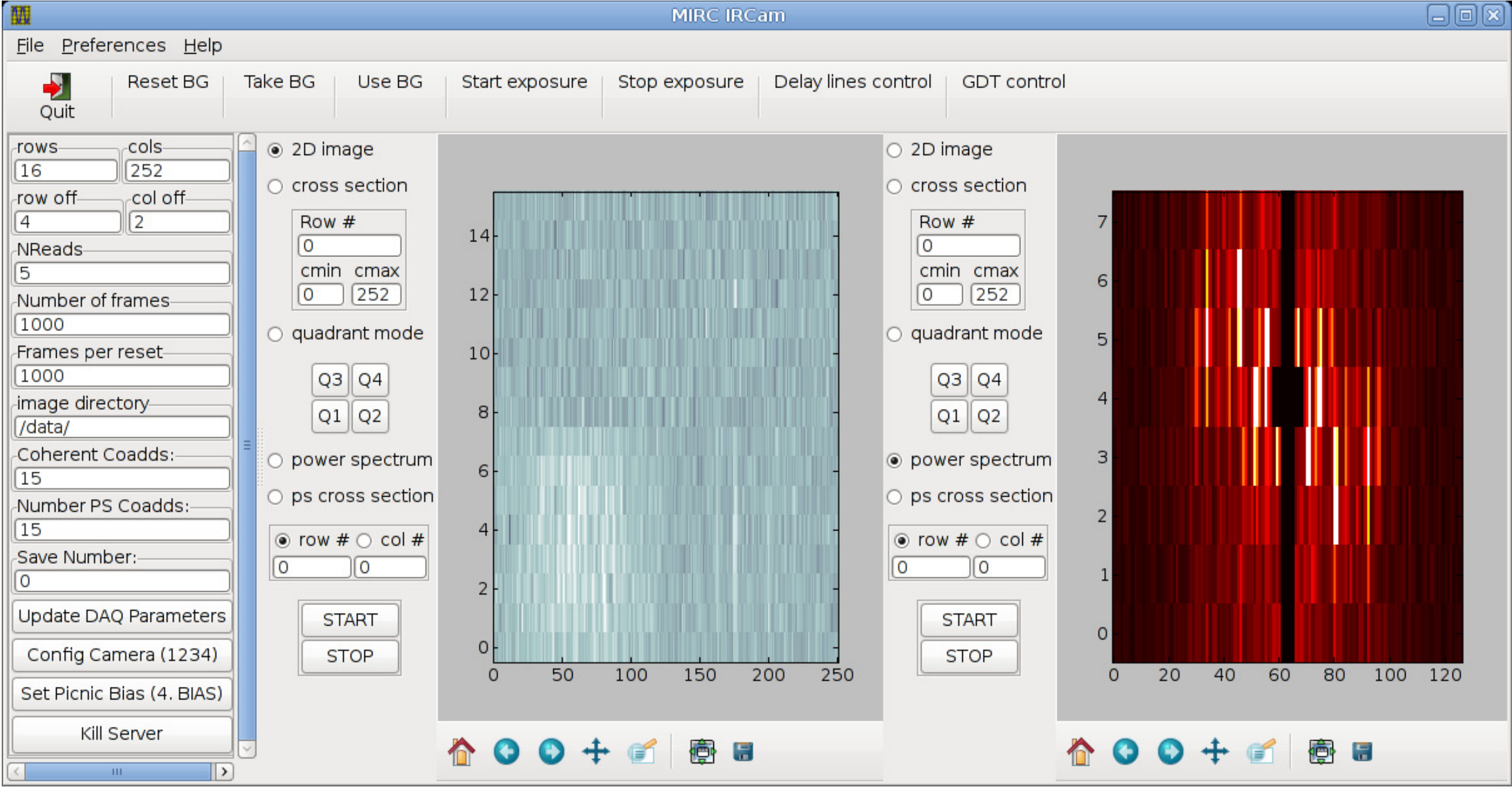}\\
  \caption{This GUI displays the MIRC real-time detector data and
    enables the user to start the fringe recording sequence.}
  \label{fig4}
\end{figure}

This GUI provides a real-time display of the detector image,
either in image space or Fourier space (Fig.~\ref{fig4}).
The left corner of the interface allows the user to change the
detector readout parameters and to start a data recording sequence by
inputting the number of requested frames into the ``Save Number'' field.

\section{CONCLUSIONS}
\label{sec:conclusions}

Increasing the number of combined apertures is essential to improve
the imaging capabilities in optical interferometry, but also poses new
challenges on operational aspects.
In this contribution, we have reviewed new control concepts that we
have developed to enable efficient instrument alignment, fringe
search, and fringe tracking with the six-telescope instruments MIRC 
and CHAMP.
Some of the discussed solutions are linked to the specific instrument
design, while others are applicable to optical interferometers in general.
For instance, the presented matrix approach constitutes a
straightforward method to utilize the cross-fringe information for
fringe search \& tracking and to display this information to the user.
Developing and optimizing such strategies is 
essential in order to pave the way for even larger optical
interferometric arrays in the future.

\acknowledgments

We thank our colleagues from the CHARA array for support in the
commissioning and operation of MIRC-6T and CHAMP.
This work was performed in part under contract with the California
Institute of Technology (Caltech) funded by NASA through the Sagan
Fellowship Program.



\bibliography{CHAMP}   
\bibliographystyle{spiebib}   

\end{document}